\documentclass[11pt]{article}

\usepackage{amsmath}
\usepackage{graphicx}
\usepackage{amsfonts}
\usepackage{amssymb}
\usepackage{epsfig}
\usepackage{color}
\usepackage{psfrag}
\usepackage{epstopdf}

\newcommand{\EQ}{\begin{equation}}
\newcommand{\EN}{\end{equation}}
\newcommand{\be}{\begin{equation}}
\newcommand{\ee}{\end{equation}}
\newcommand{\bea}{\begin{eqnarray}}
\newcommand{\eea}{\end{eqnarray}}

\begin{document} \setcounter{page}{0}
\topmargin 0pt
\oddsidemargin 5mm
\renewcommand{\thefootnote}{\arabic{footnote}}
\newpage
\setcounter{page}{0}
\topmargin 0pt
\oddsidemargin 5mm
\renewcommand{\thefootnote}{\arabic{footnote}}
\newpage
\begin{titlepage}
\begin{flushright}
%SISSA 02/2014/FISI \\
%DFTT 9/2007
\end{flushright}
\vspace{0.5cm}
\begin{center}
{\large {\bf Quantum quenches with integrable pre-quench dynamics}}\\
%{\bf Exact results from field theory}}\\
\vspace{1.8cm}
{\large Gesualdo Delfino}\\
\vspace{0.5cm}
{\em SISSA -- Via Bonomea 265, 34136 Trieste, Italy}\\
{\em INFN sezione di Trieste}\\
%{\em E-mail: delfino@sissa.it}\\
%\vspace{0.5cm}
%{\large and}\\
%\vspace{0.5cm}
%{\large P. Simonetti}\\
%\vspace{0.5cm}
%{\em Department of Physics, University of Wales Swansea,\\
%Singleton Park, Swansea SA2 8PP, United Kingdom}\\
%{\em email: p.simonetti@swansea.ac.uk}\\
\end{center}
\vspace{1.2cm}

\renewcommand{\thefootnote}{\arabic{footnote}}
\setcounter{footnote}{0}

\begin{abstract}
\noindent
We consider the unitary time evolution of a one-dimensional quantum system which is in a stationary state for negative times and then undergoes a sudden change (quench) of a parameter of its Hamiltonian at $t=0$. For systems possessing a continuum limit described by a massive quantum field theory we investigate in general perturbative quenches for the case in which the theory is integrable before the quench.
\end{abstract}
\end{titlepage}

\newpage
\noindent
Remarkable advances in experiments with cold atomic gases have triggered in the last years a strong interest in non-trivial unitary time evolution of closed quantum systems (see e.g. \cite{PSSV} for a review). A favorite case study is the quantum quench \cite{CC}, in which an eigenstate (normally the ground state) of a closed system with Hamiltonian $H_0$ evolves unitarily for times $t>0$ according to a new Hamiltonian obtained by a sudden change of a parameter of $H_0$ at $t=0$. In particular, integrable one-dimensional systems have attracted special interest due to experimental observations \cite{KWW} and theoretical proposals \cite{RDYO} suggesting that integrability or quasi-integrability may affect in a distinctive way the long time behaviour of the system. 

In this paper we consider quenches of one-dimensional quantum systems admitting a continuum limit described by a massive quantum field theory which is integrable before the quench. The theory is specified by the action
\EQ
{\cal A}={\cal A}_0-\lambda\int_0^\infty dt\int_{-\infty}^{\infty} dx\,\Psi(x,t)\,,
\label{action}
\EN
where ${\cal A}_0$ is the action of the integrable pre-quench theory, $\lambda$ the quench parameter, and $\Psi$ its conjugated local operator, which is relevant in the renormalization group sense, i.e. has scaling dimension\footnote{The theory with action ${\cal A}_0$ is an off-critical deformation by relevant operators of the conformal field theory describing a quantum critical point. All scaling dimensions refer to this critical point.} $X_\Psi<2$.

Integrability in (1+1)-dimensional quantum field theory follows from the presence of infinitely many conserved quantities and amounts to two drastic simplifications of the scattering theory \cite{ZZ}, namely complete elasticity (the final state is kinematically identical to the initial one) and complete factorization (the scattering amplitudes of $n$-particle processes factorize into the product 
of $n(n-1)/2$ two-body amplitudes). Factorization follows from the fact that conserved quantities other than energy and momentum act as generators of momentum dependent space-time translations of the trajectories of the relativistic particles, allowing to resolve a multi-particle scattering into a sequence of two-particle scatterings separated by arbitrary large distances in space-time. Once added to the general properties of unitarity and crossing symmetry, elasticity and factorization allow the exact determination of the two-particle amplitudes\footnote{In order to simplify the notation, we develop the main discussion referring to theories with a single species of particles. Generalizations will be discussed when relevant.} $S(p_1,p_2)$, and then of the whole $S$-matrix. The theory with action ${\cal A}_0$ is integrable in this sense. 

\begin{figure}[t]
\begin{center}
\includegraphics[width=13cm]{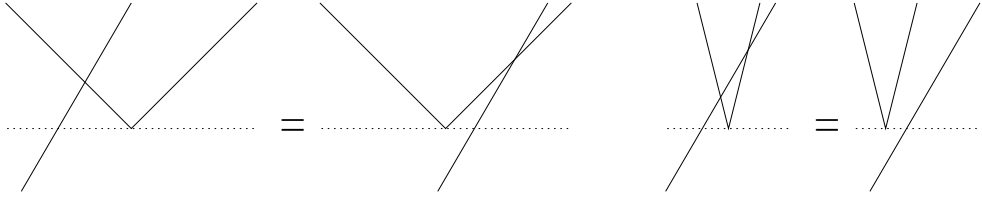}
\caption{A particle is transmitted and a pair is created at $t=0$. Pictorial illutstration of the equations following from the requirement of complete factorizabililty of the amplitude.}
\label{factorization}
\end{center}
\end{figure}

We examine first of all under which conditions a similar notion of exact solvability can extend to the quenched theory (\ref{action}). A necessary condition is that the scattering is elastic and factorized also after the quench. Concerning the passage from negative to positive times, requiring elasticity means that particles transmitted through $t=0$ preserve their momentum; on the other hand, the breaking of time translation invariance at $t=0$ allows for creation or destruction of sets of particles with zero total momemtum (momentum is conserved by (\ref{action})). A typical example of a process in the quenched theory is that in which a particle of momentum $p$ is transmitted through $t=0$ and scatters with a pair of particles with momenta $q$ and $-q$ produced at $t=0$. The requirement of complete factorizability implies that the total amplitude must be unaffected by an arbitrary translation of, say, the trajectory of the transmitted particle. Depending on the value of $q$ this gives rise to two equations that are depicted in Fig.~\ref{factorization}. The total amplitudes contain common factors $T(p)$ and $K(q)$, namely the transmission amplitude for the particle with momentum $p$ and the creation amplitude for the pair, respectively. Factorizability then constrains the post-quench amplitudes $S(p_1,p_2)={\cal S}(\theta_1-\theta_2)$, where we introduced the rapidity parameterization $(E_p,p)=(m\cosh\theta,m\sinh\theta)$ for the energy and momentum of a relativistic particle with mass $m$; the dependence of the two-particle amplitude on the rapidity difference follows from relativistic invariance. The pictorial equations of Fig.~\ref{factorization} correspond to
\bea
& {\cal S}(\theta+\beta)={\cal S}(\beta-\theta)\,,\nonumber\\
& {\cal S}(\theta+\beta){\cal S}(\theta-\beta)=1\,.\nonumber
\eea
The first equation implies that ${\cal S}$ is a constant, and the second (or, equivalently, unitarity) that ${\cal S}=\pm 1$, namely that the post-quench particles are either free bosons or free fermions\footnote{In the context of theories with a defect line a similar factorization argument was used in \cite{DMS} and generalized to more particles species in \cite{CFG}.}. We arrive at the same conclusion for the pre-quench particles if  we consider the process in which an initial state with three particles with momenta $p$, $q$ and $-q$ evolves into the final state with a single particle with momentum $p$ through pair desctruction at $t=0$. The conclusion we reach in this way is that the only quenches compatible with factorization are mass quenches for free particles (i.e. ${\cal A}_0$ is a free action and $\lambda\Psi$ is proportional to its mass term). The amplitudes $T(p)$ and $K(p)$ have been determined by Bogoliubov transformations in \cite{CC} for free bosons and in \cite{RSMS} for free fermions.

This argument indicates that the only way to deal analytically with interacting particles in quenched massive field theories is the perturbative one. Here we treat (\ref{action}) perturbatively in $\lambda$, keeping ${\cal A}_0$ integrable. Perturbation theory around integrable quantum field theories in the time translation invariant case was formulated in \cite{nonint}. In the present case the perturbation $\lambda\Psi$ acts only for positive times, but it is still possible to set up a scattering description based on the {\em asymptotic states} $|p_1,\ldots,p_n\rangle_{in\,(out)}$ {\em of the integrable pre-quench theory}. An initial state $|p_1,\ldots,p_n\rangle_{in}$ (which is an eigenstate of the pre-quench Hamiltonian $H_0$ with eigenvalue $\sum_{i=1}^nE_{p_i}$, $E_{p_i}=\sqrt{m^2+p_i^2}$) at $t=-\infty$ evolves at $t=+\infty$ into a final state that we expand on the complete basis of outgoing states of the pre-quench theory. The coefficients of this expansion are
\EQ
%S(p_1,\ldots,p_n|q_1,\ldots,q_m)=
{}_{out}\langle q_1,\ldots,q_m|S_\lambda|p_1,\ldots,p_n\rangle_{in}\,,
\label{coeff}
\EN
with 
\EQ
S_\lambda=T\,\exp\left(-i\lambda\int_0^\infty dt\int_{-\infty}^\infty dx\,\Psi(x,t)\right)
\label{Slambda}
\EN
($T$ indicates chronological ordering); $S_\lambda$ becomes the identity for $\lambda=0$ and (\ref{coeff}) reduces to the elastic and factorized $S$-matrix of the integrable theory with action ${\cal A}_0$. 

We now consider the case in which the system is for $t<0$ in the ground state (i.e. in the vacuum state $|0\rangle$) and $\lambda$ is a small parameter. Then, to lowest order in $\lambda$, the state $|\psi_0\rangle$ of the system after the quench can be written as
\EQ
|\psi_0\rangle=S_\lambda|0\rangle\simeq |0\rangle+\lambda\sum_{n=1}^\infty\frac{2\pi}{n!}\int_{-\infty}^{\infty}\prod_{i=1}^n\frac{dp_i}{2\pi E_{p_i}}\,\delta(\sum_{i=1}^np_i)\,\frac{[F_n^\Psi(p_1,\ldots,p_n)]^*}{\sum_{i=1}^nE_{p_i}}\,|p_1,\ldots,p_n\rangle\,,
\label{psi0}
\EN
\EQ
F_n^\Psi(p_1,\ldots,p_n)=\langle 0|\Psi(0,0)|p_1,\ldots,p_n\rangle\,.
\label{ff}
\EN
The matrix elements (\ref{ff}) are the form factors of the integrable theory with action ${\cal A}_0$ and are exactly computable (see \cite{Smirnov}). The factor $\sum_{i=1}^nE_{p_i}$ in the denominator of (\ref{psi0}) is obtained giving to the energy an infinitesimal imaginary part to make the time integral in (\ref{Slambda}) convergent. In writing (\ref{psi0}) we omitted the $n=0$ term of the sum, which reads $\lambda LT_+\langle 0|\Psi|0\rangle|0\rangle$, where $LT_+$ is the (infinite) post-quench space-time volume; this term corresponds to a renormalization of the vacuum energy density and is canceled introducing a corresponding counterterm in the exponent of (\ref{Slambda}) (see \cite{nonint} for details on an analogous procedure). In (\ref{psi0}) we also passed from the outgoing states $|p_1,\ldots,p_n\rangle_{out}$ and the corresponding integration over momenta restricted by $p_1<p_2<\ldots<p_n$ to the generic states $|p_1,\ldots,p_n\rangle$ and $1/n!$ times unrestricted integration. This is possible because interchanging the position of the particles with momenta $p_i$ and $p_{i+1}$ yields a factor\footnote{Here $S(p_1,p_2)$ denotes the scattering amplitude in the integrable pre-quench theory.} $S(p_i,p_{i+1})$ from $|p_1,\ldots,p_n\rangle$, together with a factor $S^*(p_i,p_{i+1})$ from the form factor, and the two factors cancel by unitarity. Since to lowest order the particles scatter after the quench with the same factorized pre-quench $S$-matrix, the independence on the ordering of the particles means that (\ref{psi0}) describes the state of the system at any positive time, i.e. it can be considered as a state of the Heisenberg picture in which all the time dependence is carried by the operators. 

The $n=2$ term of (\ref{psi0}) reads $\frac{1}{2}\int\frac{dp}{2\pi E_p}K^{(1)}(p)|p,-p\rangle$, with $K^{(1)}(p)=\lambda\frac{[F_2^\Psi(p,-p)]^*}{2E^2_p}$. For the mass quench in free theories one can verify that $K^{(1)}(p)$ coincides with the lowest order of the pair creation amplitude obtained by Bogoliubov transformation. In these solvable cases $F_n^\Psi$ vanishes for $n\neq 2$.

The average energy produced in the quench is\footnote{When computing averages on $|\psi_0\rangle$ we should divide by $\langle\psi_0|\psi_0\rangle$. This normalization, however, is $1+O(\lambda^2)$ and can be ignored to lowest order.}
\EQ
\langle\psi_0|H_0|\psi_0\rangle\simeq\lambda^2\,L\sum_{n=1}^\infty\frac{2\pi}{n!}\int_{-\infty}^{\infty}\prod_{i=1}^n\frac{dp_i}{2\pi E_{p_i}}\,\delta(\sum_{i=1}^np_i)\,\frac{|F_n^\Psi(p_1,\ldots,p_n)|^2}{\sum_{i=1}^nE_{p_i}}\,,
\label{energy}
\EN
where $L=2\pi\delta(0)\to\infty$ is the size of system. Since it was shown in \cite{immf} that for a scalar operator $\Phi$ with scaling dimension $X_\Phi$
\EQ
F_n^\Phi(p_1,\ldots,p_n)\sim |p_i|^{y_\Phi}\,,\hspace{.6cm}|p_i|\to\infty\,;\hspace{1.2cm}y_\Phi\leq X_\Phi/2\,,
\label{asymp}
\EN
a sufficient condition for the convergence of the integrals in (\ref{energy}) is\footnote{If this condition is not satisfied one can resort to a more specific asymptotic analysis \cite{D_09}. Divergent integrals at this stage signal operator mixing; see \cite{DSC} for a discussion in the context of a different application.} $X_\Psi<1$. Typically the contribution of a state to a form factor expansion rapidly decreases as the total mass increases.

Using the notation $\delta F(\lambda)=F(\lambda)-F(0)$, we have for one-point functions of hermitian operators
\bea
\delta\langle\psi_0|\Phi(x,t)|\psi_0\rangle &\simeq &\lambda\sum_{n=1}^\infty\frac{2\pi}{n!}\int_{-\infty}^{\infty}\prod_{j=1}^n\frac{dp_j}{2\pi E_{p_j}}\,\frac{\delta(\sum_{j=1}^np_j)}{\sum_{j=1}^nE_{p_j}}\,\nonumber\\
&\times& 2\mbox{Re}\{[F_n^\Psi(p_1,\ldots,p_n)]^*F_n^\Phi(p_1,\ldots,p_n)\,e^{-i\sum_{j=1}^nE_{p_j}t}\}\,,
\label{1point}
\eea
where the integrals converge for $\Phi$ relevant.

When considering the long time behavior we have to take into account that there are two time scales, i.e. $1/m$ and $t_\lambda=1/\lambda^{1/(2-X_\Psi)}$; the latter goes to infinity as $\lambda$ goes to zero. Within the perturbative approach in $\lambda$, long time means staying in the intermediate regime $1/m\ll t\ll t_\lambda$, so that the limit of small $\lambda$ prevails. At large times the integrand of (\ref{1point}) rapidly oscillates, we  replace $e^{-iE_{p_j}t}$ with $e^{-i(m+p_j^2/2m)t}$, and the integrals receive the main contribution from a range of $|p_j|$ with width of order $\sqrt{m/t}$. Since, with the exception of free bosons, the scattering amplitude of two identical particles satisfies $S(p,p)=-1$, the factor $[F_n^\Psi]^*F_n^\Phi$ in (\ref{1point}) evaluated for momenta all tending to zero becomes $\prod_{1\leq i<k\leq n}(p_i-p_k)^2$ times a constant. Hence, following this analysis, the passage to the variables $q_j=\sqrt{\frac{t}{m}}p_j$ gives for the $n$-particle term in (\ref{1point}) a leading intermediate time behavior
\EQ
\frac{\lambda}{t^{(n^2-1)/2}}\,\mbox{Re}(c_n\,e^{-inmt})\,,
\label{leading}
\EN
with coefficients $c_n$ which depend on ${\cal A}_0$, $\Psi$ and $\Phi$. In theories with more particle species, particles of different species with momenta $p_i$ and $p_k$ do not necessarily contribute to the integrand the factor $(p_i-p_k)^2$, and time suppression for $n>1$ may be reduced. In all cases, however, if the form factors\footnote{One-particle form factors of scalar operators are momentum-independent.} $F_{1,a}^\Psi$ and $F_{1,a}^\Phi$ over a single particle of species $a$ are both non-zero, the leading intermediate time behavior is  
\EQ
\lambda\sum_a\frac{2}{m_a^2}\mbox{Re}\{[F_{1,a}^\Psi]^*F_{1,a}^\Phi\,e^{-im_at}\}\,.
\label{leading1}
\EN

A convenient illustration is provided by the Ising field theory (see \cite{review} for a review), namely the theory describing the scaling region around the simplest quantum critical point possessing a $Z_2$ symmetry. With reference to the action (\ref{action}), we consider first the case in which ${\cal A}_0$ is massive and $Z_2$-symmetric; this is the continuum limit of the transverse field Ising spin chain and corresponds to the theory of a free neutral fermion. The first possibility is to take $\Psi$ equal to $\varepsilon$, the relevant $Z_2$-invariant operator ($X_\varepsilon=1$), which with the present choice of ${\cal A}_0$ has $F_n^\varepsilon\neq 0$ only for $n=2$. This quench corresponds to one of the exactly solvable cases (mass quench for free fermions, $\lambda=\delta m$). The intermediate time behavior of (\ref{1point}) for the order parameter operator $\sigma$ ($X_\sigma=1/8$) in the ferromagnetic phase (the one-point function vanishes by symmetry in the paramagnetic phase) corresponds to (\ref{leading}) with $n=2$; this agrees with the result of \cite{CEF,SE}. The second possibility is to take $\Psi=\sigma$, a choice which breaks $Z_2$ symmetry after the quench. Now the intermediate time behavior of $\delta\langle\psi_0|\sigma(x,t)|\psi_0\rangle$ should be given by (\ref{leading}) with $n=1$ if we start from the paramagnetic phase, and $n=2$ if we start from the ferromagnetic phase\footnote{The operator $\sigma$ has non-zero form factors on states with odd (resp. even) number of particles in the paramagnetic (resp. ferromagnetic) phase.}.

If instead we move away from the quantum critical point in the direction of the operator $\sigma$ we obtain a theory ${\cal A}_0$ which is no longer $Z_2$ invariant but is still integrable and contains eight species of particles with different masses \cite{Taniguchi} (see \cite{Coldea} for an experimental study). Both for $\Psi=\sigma$ and $\Psi=\varepsilon$ the leading intermediate time behavior of $\delta\langle\psi_0|\sigma(x,t)|\psi_0\rangle$ is given by (\ref{leading1}), with form factors $F_{1,a}^\sigma$ and $F_{1,a}^\varepsilon$ ($a=1,\ldots,8$) which are also known \cite{immf,DS,review}. The case $\Psi=\sigma$ now corresponds to the quench of the mass scale in an interacting theory.

Two-point correlators $\langle\psi_0|\Phi_1(x_1,t_1)\Phi_2(x_2,t_2)|\psi_0\rangle$ can also be expanded over form factors inserting a complete set of states of the pre-quench theory in between the two operators. This results in a double sum over particle states that we do not try to analyze here.

\vspace{.3cm}
Summarizing, we considered quantum quenches in massive (1+1)-dimensional field theory. After arguing that exactly solvable cases restrict to mass quenches in free theories, we formulated the perturbative theory in the quench parameter for the case of integrable pre-quench dynamics. In particular, we determined to lowest order the average energy per unit length produced in the quench and analyzed the behavior of one-point functions of local operators for times much larger than the pre-quench time scale but much smaller than the time scale associated to the quench parameter.

%\newpage

%\vspace{1cm} \noindent \textbf{Acknowledgments.} 


\begin{thebibliography}{99}
\bibitem{PSSV} A. Polkovnikov, K. Sengupta, A. Silva and M. Vengalattore, Rev. Mod. Phys. 83 (2011) 863.
\bibitem{CC} P. Calabrese and J. Cardy, Phys. Rev. Lett. 96 (2006) 136801; J. Stat. Mech. (2007) P06008.
\bibitem{KWW} T. Kinoshita, T. Wenger and D.S. Weiss, Nature 440 (2006) 900.
\bibitem{RDYO} M. Rigol, V. Dunjko, V. Yurovsky and M. Olshanii, Phys. Rev. Lett. 98 (2007) 50405.
\bibitem{ZZ} A.B. Zamolodchikov and Al.B. Zamolodchikov, Ann. Phys. 120 (1979) 253.
\bibitem{RSMS} D. Rossini, A. Silva, G. Mussardo and G. Santoro, Phys. Rev. Lett. 102 (2009) 127204.
\bibitem{DMS} G. Delfino, G. Mussardo and P. Simonetti, Phys. Lett. B 328 (1994) 123; Nucl. Phys. B 432 (1994) 518.
\bibitem{CFG} O. Castro-Alvaredo, A. Fring and F. Gohmann, arXiv:hep-th/0201142.
\bibitem{nonint} G. Delfino, G. Mussardo and P. Simonetti, Nucl. Phys. B 473 (1996) 469. 
\bibitem{Smirnov} F.A. Smirnov, Form factors in completely integrable models of quantum field theory, World Scientific, Singapore, 1992.
\bibitem{immf} G. Delfino and G. Mussardo, Nucl. Phys. B 455 (1995) 724.
\bibitem{D_09} G. Delfino, Nucl. Phys. B 807 (2009) 455.
\bibitem{DSC} G. Delfino, P. Simonetti and J. Cardy, Phys. Lett. B 387 (1996) 327.
\bibitem{review} G. Delfino, J. Phys. A 37 (2004) R45.
\bibitem{CEF} P. Calabrese, F.H.L. Essler and M. Fagotti, Phys. Rev. Lett.
106 (2011) 227203; J. Stat. Mech. (2012) P07016.
\bibitem{SE} D. Schuricht and F. Essler, J. Stat. Mech. (2012) P04017.
\bibitem{Taniguchi} A.B. Zamolodchikov, Int. J. Mod. Phys. A 4 (1989) 4235.
\bibitem{Coldea} R. Coldea, D.A. Tennant, E.M. Wheeler, E. Wawrzynska, D.
Prabhakaran, M. Telling, K. Habicht, P. Smeibidl and K. Kiefer, Science 327 (2010) 177.
\bibitem{DS} G. Delfino and P. Simonetti, Phys. Lett. B 383 (1996) 450.


\end{thebibliography}
\end{document}